\documentclass[journal]{IEEEtran}
\usepackage{hyperref}
\usepackage[T1]{fontenc}
\usepackage{subfigure}
\usepackage[cmex10]{amsmath}
\usepackage{amsmath}
\usepackage{amssymb}
\usepackage{multirow}
\usepackage[cmex10]{amsmath}
\usepackage{cases}
\usepackage{soul}
\usepackage{threeparttable}
\usepackage{color}
\usepackage{flushend} 
\ifCLASSINFOpdf
  \usepackage[pdftex]{graphicx}
  \graphicspath{{../pdf/}{../jpeg/}}
  \DeclareGraphicsExtensions{.pdf,.jpeg,.png}
\else
  \usepackage[dvips]{graphicx}
  \graphicspath{{../eps/}}
  \DeclareGraphicsExtensions{.eps}
\fi

\newcommand{\ms}{\scriptscriptstyle}
\newcommand{\sinc}{\mathrm{sinc} \,}
\graphicspath{{./Figures/}}

\graphicspath{{./Figures/}}
\hypersetup
{
    pdftitle={Correlated_Photon_Pairs_2011},
    pdfauthor={\'Alvaro Almeida},
    colorlinks=black,
    linkcolor=black,
    citecolor=black,
    pdfview=Fit,
    pdfpagemode=UseNone
}

\begin{document}

\title{Correlated Photon-Pair Generation in a Highly Nonlinear Fiber Using Spontaneous FWM}
\author{Álvaro~J.~Almeida, Lu\'is M. Martins, Nuno~A.~Silva, Nelson~J.~Muga, and Armando~N.~Pinto
\thanks{This work was partially supported by the Funda\c{c}\~{a}o para a Ci\^{e}ncia e Tecnologia, FCT, through the Laborat\'orio Associado (IT/LA) program, project ``QuantPrivTel - Quantum Private Telecommunications'' project (PTDC/EEA-TEL/103402/2008), FEDER and PTDC program.}

\thanks{Álvaro~J.~Almeida is with the Instituto de Telecomunica\c{c}\~{o}es, University of Aveiro, Campus Universit\'{a}rio de Santiago, 3810-193 Aveiro, Portugal, Tel: +351 234 377 900, Fax: +351 234 377 901 (e-mail: aalmeida@av.it.pt).}

\thanks{Lu\'is~M.~Martins and Nelson~J.~Muga are with the Department of Physics, University of Aveiro, and the Instituto de Telecomunica\c{c}\~{o}es, University of Aveiro, Campus Universit\'{a}rio de Santiago, 3810-193 Aveiro, Portugal, Tel: +351 234 377 900, Fax: +351 234 377 901 (e-mails: lmartins@av.it.pt; muga@av.it.pt).}

\thanks{Nuno~A.~Silva and Armando~N.~Pinto are with the Department de Electronics, Telecommunications and Informatics, University of Aveiro, and the Instituto de Telecomunica\c{c}\~{o}es, University of Aveiro, Campus Universit\'{a}rio de Santiago, 3810-193 Aveiro, Portugal, Tel: +351 234 377 900, Fax: +351 234 377 901 (e-mails: nasilva@av.it.pt; anp@ua.pt).}}
\maketitle
\thispagestyle{empty} 
\pagestyle{empty} 

\begin{abstract}
We generate 1550~nm correlated photon pairs through the spontaneous four-wave mixing (SpFWM) process in a highly nonlinear fiber (HNLF).
The pair source quality generation is evaluated by the coincidence-to-accidental ratio (CAR) parameter. It is verified that the spontaneous Raman scattering (SpRS) photons generated inside the fiber contribute to the degradation of the source. Nevertheless, a CAR value of 2.77 is found for a pump power at the input of the fiber of 1.1~mW.
\end{abstract}

\begin{IEEEkeywords}
Spontaneous Four-Wave Mixing, Spontaneous Raman Scattering, Coincidence-to-Accidental Ratio, Quantum Correlation, Quantum Communication.
\end{IEEEkeywords}

\section{Introduction}

\IEEEPARstart{T}{he} generation of 1550~nm correlated photon pairs has been widely studied in the last few years. In optical fibers, this has been achieved for example via the four-wave mixing (FWM) process~\cite{Wang.2001JOptB...3..346W,Inoue.2004JaJAP..43.8048I,Takesue.2005OExpr..13.7832T}.
In the same way, correlated photons have been generated from photonic-chip platforms based on parametric downconversion in periodically poled lithium niobate (PPLN) waveguides~\cite{2010OptL...35.1239H}, and from spontaneous four-wave mixing (SpFWM) in silicon nanowires~\cite{2006OExpr..1412388S,2010OptL...35.3483C,Harada.2010}. Recently, the generation of 1550~nm correlated photon pairs by SpFWM in an integrated chalcogenide As$_2$S$_3$ waveguide, using a continuous wave (cw) pump, was reported in~\cite{Xiong.2011ApPhL..98e1101X}. These structures present some peculiar advantages, such as an high third-order nonlinearity~\cite{Ta'Eed.2007OExpr..15.9205T}, or an high parametric gain~\cite{Lamont.2008OExpr..1620374L}. However, build them with some specific characteristics is not yet an easy task~\cite{Xiong.2011ApPhL..98e1101X}.

In this paper, we investigate both theoretically and experimentally the generation of correlated photon pairs in a highly nonlinear fiber (HNLF), in the 1550~nm wavelength band. It is well known that in standard dispersion-shifted fibers (DSFs), together with the photon-pair generation through the SpFWM process, noise photons are generated by spontaneous Raman scattering (SpRS). These photons lead to a significant deterioration on the performance of the correlated photon-pair generation. Cooling the fiber is a way to reduce significantly the noise photons generation~\cite{Takesue.2005OExpr..13.7832T}. When compared to DSFs, HNLFs presents an higher nonlinear parameter that leads also to a lower rate of noise-photon generation, since lower powers can be used~\cite{Zhou:10}. In this way, the use of HNLFs brings a new window of opportunities to this field, which can be extensively explored. In order to evaluate the correlated photon-pair generation in our source, we use the ratio between coincident and accidental counts, that is commonly referred as coincidence-to-accidental ratio (CAR).

This paper contains five sections. In Section~\ref{td}, we present a theoretical formulation for the generation of correlated photon pairs by SpFWM, and noise photons from SpRS. The model for single-, coincidence- and accidental-count rates detection is also presented. In Section~\ref{ed}, the experimental setup used in our experiment is described. In Section~\ref{er}, a comparison between the experimental and the theoretical results is presented. The main conclusions of this paper are presented in Section~\ref{c}.

\section{Theoretical Description}
\label{td}

A correlated photon-pair source outputs signal-idler photon pairs, which exhibit a temporal correlation. In another way, it is a source that generates a product state, instead of an entangled state~\cite{Fiorentino.2002IPTL...14..983F,Takesue.2005OExpr..13.7832T,AA.2011.EURCON}. The photon pair production rate by SpFWM is given by~\cite{Zhou:10,Xiong.2011ApPhL..98e1101X},
\begin{align}
R_{\ms \textrm{SpFWM}} = \sigma \Delta\nu |\gamma P_0 L|^2 \sinc^2\bigg[\bigg((2\pi\nu)^2\beta_2 + 2\gamma P_0\bigg) \frac{L}{2} \bigg]\,.
\label{RSpFWM}
\end{align}
In~\eqref{RSpFWM}, $\sigma = \tau f_\textrm{p}$, is a duty cycle parameter, where $\tau$ is the pulse full width at half maximum (FWHM), and $f_\textrm{p}$ is the pulse repetition rate, and $\Delta\nu$ represents the filter bandwidth. The frequency detuning, $\nu = (\omega_\textrm{s}-\omega_\textrm{p})/2\pi = (\omega_\textrm{p}-\omega_\textrm{i})/2\pi$, and $\omega_\textrm{p}$, $\omega_\textrm{s}$ and $\omega_\textrm{i}$ are the pump, signal and idler frequencies, respectively. The parameters $\gamma$, $P_0$ and $L$ are the nonlinear coefficient, the input pump power and the length of the fiber, respectively, and $\beta_2 = d^2\beta/d\omega^2$ is the group velocity dispersion parameter, measured at $\omega_\textrm{p}$.

The noise photon generation through SpRS in the signal channel is written as~\cite{Zhou:10,Xiong.2011ApPhL..98e1101X},
\begin{align}
R_{\ms \textrm{SpRS}}^{(s)} = \sigma \Delta\nu P_0 L |g_\textrm{R}(\nu_\textrm{s})|(n_{\textrm{th}}^{s})\,,
\label{RSpRSs}
\end{align}
and in the idler channel, as,
\begin{align}
R_{\ms \textrm{SpRS}}^{(i)} = \sigma \Delta\nu P_0 L |g_\textrm{R}(\nu_\textrm{i})|(n_{\textrm{th}}^{i}+1)\,,
\label{RSpRSi}
\end{align}
for the case when $\lambda_s<\lambda_p<\lambda_i$. In~\eqref{RSpRSs} and~\eqref{RSpRSi}, the Raman gain spectrum, $g_\textrm{R}(\nu_{\textrm{s,i}})$, is defined as,
\begin{align}
g_{\ms \textrm{R}}(\nu_{\textrm{s,i}}) = 2\gamma f_\textrm{R} \textrm{Im}[h_\textrm{R}(\nu_{\textrm{s,i}})]\,,
\label{gr}
\end{align}
where $f_\textrm{R}$ is the fractional contribution of the delayed Raman response, and $h_{\ms \textrm{R}}(\nu_{\textrm{s,i}})$ is the Raman response function for the signal and the idler, respectively~\cite{2006nfo..book.....A}.
The parameter $n_{\textrm{th}}^{s,i}$ is the phonon population at frequency $\nu_{\textrm{s,i}}$ and temperature $T$, and is described by the Bose-Einstein distribution,
\begin{align}
n_{\textrm{th}}^{s,i} = \frac{1}{\exp(h\nu_{\textrm{s,i}}/k_\textrm{B} T)-1}\,,
\end{align}
where $h$ and $k_\textrm{B}$ are, respectively, Planck's and Boltzmann's constants.

The rate of single counts at the two detectors, respectively from the signal and the idler photons, is given by~\cite{Zhou:10,Xiong.2011ApPhL..98e1101X},
\begin{align}
N_\textrm{s} = \eta_\textrm{s}(R_{\ms \textrm{SpFWM}}+R_{\ms \textrm{SpRS}}^{(s)}) + d_\textrm{s}\,,
\label{ns}
\end{align}
and,
\begin{align}
N_\textrm{i} = \eta_\textrm{i}(R_{\ms \textrm{SpFWM}}+R_{\ms \textrm{SpRS}}^{(i)}) + d_\textrm{i}\,.
\label{ni}
\end{align}
The parameters $\eta_{\textrm{s,i}}$ represents the system detection efficiencies of the signal and idler paths, respectively, and these include the avalanche photodetectors (APDs) quantum efficiencies, the coupling light losses from the fiber to the detectors, and the insertion losses of the signal and idler filters. The parameters $d_{\textrm{s,i}}$ are the dark-count rates of the APDs at the signal and idler sides.

The coincidence-count rate is given by~\cite{Fan.2005OptL...30.1530F,Zhou:10},
\begin{align}
N_{\textrm{co}} = \eta_{\ms \textrm{$\alpha$}}^2(\eta_\textrm{s}\eta_\textrm{i}R_{\ms \textrm{SpFWM}} + \frac{N_\textrm{s}N_\textrm{i}}{f_\textrm{p}})\,,
\label{nco}
\end{align}
and the accidental-count rate can be written as~\cite{Fan.2005OptL...30.1530F,Zhou:10},
\begin{align}
N_{\textrm{acc}} = \eta_{\ms \textrm{$\alpha$}}^2(\frac{N_\textrm{s}N_\textrm{i}}{f_\textrm{p}})\,,
\label{nac}
\end{align}
where $\eta_{\ms \textrm{$\alpha$}}$ is the efficiency of the coincidence counter.

Finally, the CAR can be defined as,
\begin{align}
CAR = \frac{N_{\textrm{co}}}{N_{\textrm{acc}}}\,.
\label{careq}
\end{align}
The source will be as good and pure as higher the CAR parameter~\cite{2010OptCo.283..276T}.

\section{Description of the Experimental Measurement Method of Generation and Detection}
\label{ed}

\subsection{Correlated Photon-Pair Generation Through SpFWM}
\label{cppg}

\begin{figure}[!t]
  \centering
  \includegraphics[width=8.5cm]{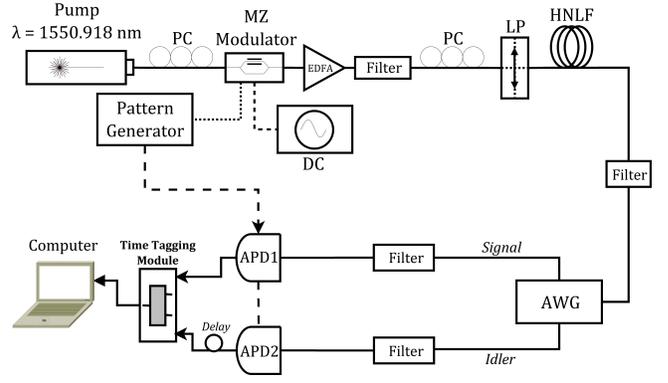}\\
  \caption{Schematics of the experimental setup used to generate correlated photon pairs from SpFWM process.}
  \label{setup}
\end{figure}

The schematics of the setup used in our experiment is presented in Fig.~\ref{setup}.
The pump is launched from a tunable laser source, that is centered at 1550.918~nm. Then, it passes through a polarization controller (PC), in order to maximize the efficiency of the Mach-Zehnder (MZ)~\cite{Muga.2006JLwT...24.3932M}. The MZ modulator is connected to a DC voltage source and to a pattern generator. At the output of the MZ, the pulse presents a FWHM of $\approx$\!1~ns and a repetition rate of 2.4~MHz. Next, the pulse intensity is amplified using an erbium doped fiber amplifier (EDFA). The sidebands of the laser and the noise introduced with the EDFA are eliminated by using a 100~GHz flat-top fixed optical filter. A second PC and a linear polarizer (LP) are used to assure that the photons are linearly polarized at the input of the HNLF. Due to the SpFWM process, a signal and an idler wave will be simultaneously generated in fiber, and are correlated in time. The pump, signal and idler photons, with frequencies $\omega_\textrm{p}$, $\omega_\textrm{s}$ and $\omega_\textrm{i}$, respectively, satisfy the phase-matching condition, given by $2\omega_\textrm{p} = \omega_\textrm{s} + \omega_\textrm{i}$~\cite{2006nfo..book.....A}.

After the HNLF, a second fixed optical filter, is used in order to suppress pump photons. Then, the idler and signal photons are separated by an arrayed waveguide grating (AWG), with a 200~GHz channel spacing. The AWG output channels with peak wavelengths of $\lambda_\textrm{s}\!=\!1547.715$~nm and $\lambda_\textrm{i}\!=\!1554.134$~nm were used for the signal and the idler, respectively. Next, a cascade of filters centered at the AWG's output channels are used in each arm, in order to assure that only signal or idler photons pass. In total, pump suffers a suppression in power of about 130~dB.

\subsection{Single and Correlated Photons Detection}
\label{scpd}

At the detection side, each photon goes into an InGaAs/InP avalanche photodetector (APD1 and APD2) from IdQuantique, operating in a gated Geiger mode~\cite{Ribordy.2004JMOp...51.1381R}. APD1 (id201) and APD2 (id200), have a dark-count probability per time gate, $t_g\!=\!5$~ns, of $P_{\textrm{dc}}\!\approx\!5\!\times\!10^{-6}$ and $P_{\textrm{dc}}\!\approx\!1\!\times\!10^{-5}$, respectively, and a quantum detection efficiency, $\eta_{\ms \textrm{D}}\!\approx\!10\%$. In order to avoid afterpulses, a 10~$\mu$s deadtime was applied to both detectors. The electric signals from the APDs were input into a time tagging module (TTM) for coincidence measurements, that worked in a continuous mode, with a time resolution of 82.3~ps.
The detector that collects the signal photons is used as a start pulse for the TTM, and the idler photons, that are delayed by using an additional electrical cable, works as a stop pulse. The reason for insert this delay was that the coincidence events caused by the photons generated at the same time, do not fall within the TTM deadtime. Thereby, we can obtain an histogram of coincidence events as a function of time. In case of existing a temporal correlation between a signal and an idler photon, a peak is observed in the histogram in the temporal position that corresponds to the \emph{coincidence} events. We also observe side peaks, which are the coincidence events induced by the photons generated at different temporal positions. These events are called \emph{accidental}~\cite{2010OptCo.283..276T}.

\section{Comparison Between Theoretical and Experimental Results}
\label{er}

\begin{figure}[!t]
\centering
\includegraphics[width=8.5cm]{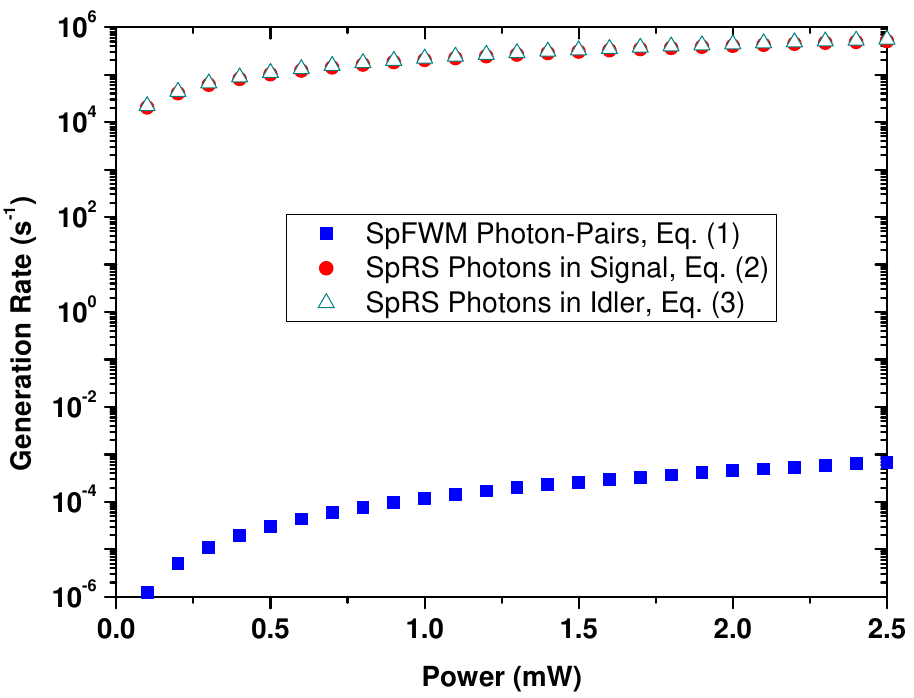}
\caption{Generation rate of photon pairs through SpFWM and noise photons from SpRS as a function of the pump power at the input of the fiber.}
\label{photonpairs}
\end{figure}

The experimental setup used to generate 1550~nm correlated photon pairs in a HNLF is presented in Fig.~\ref{setup}. The first parameter that we should evaluate is the photon-pair production rate through SpFWM process. This can be calculated according to~\eqref{RSpFWM}. In the same way, the noise photons generated from the SpRS, can be calculated using~\eqref{RSpRSs} and~\eqref{RSpRSi}. The photon-pair generation rate by SpFWM process and the noise photon generation through SpRS as a function of the pump power at the input of the fiber are plotted in Fig.~\ref{photonpairs}. The fiber parameters are the following: length, $L=$500~m, group velocity dispersion parameter, $\beta_2\!=\!-6.90\!\times\!10^{-25}$~s$^{2}$m$^{-1}$, (measured with an optical spectrum analyzer (ONA)), and a nonlinear parameter, $\gamma$~=~10.5~W$^{-1}$km$^{-1}$. We assumed that the fiber was at the room temperature, $T\!\approx\!291$~K.
\begin{figure}[!b]
\centering
\includegraphics[width=8.5cm]{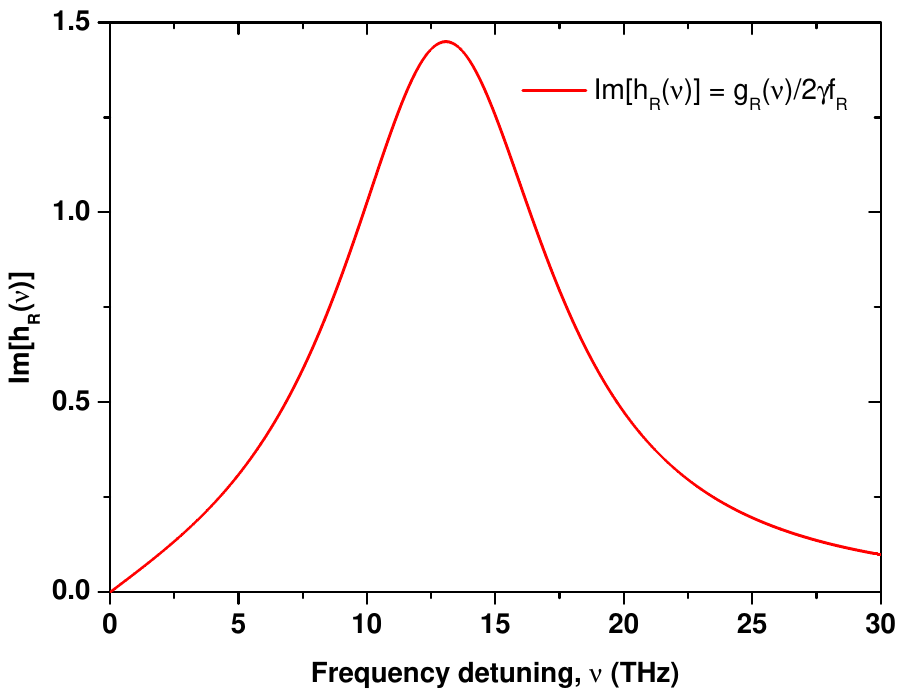}
\caption{Raman response function as a function of the frequency detuning.}
\label{raman}
\end{figure}
In order to calculate the photon generation through SpRS, both in signal and idler sides, we have calculated the Raman response function of the HNLF, $h_\textrm{R}$, as a function of the frequency detuning, $\nu$. For that purpose, we have used~\eqref{gr}, considering $f_\textrm{R} = 0.18$~\cite{2006nfo..book.....A}. The resultant spectrum of the imaginary part of the Raman response function, formulated in~\cite{2006nfo..book.....A}, for a 30~THz bandwidth, is presented in Fig.~\ref{raman}.

\begin{figure}[!t]
\centering
\includegraphics[width=8.5cm]{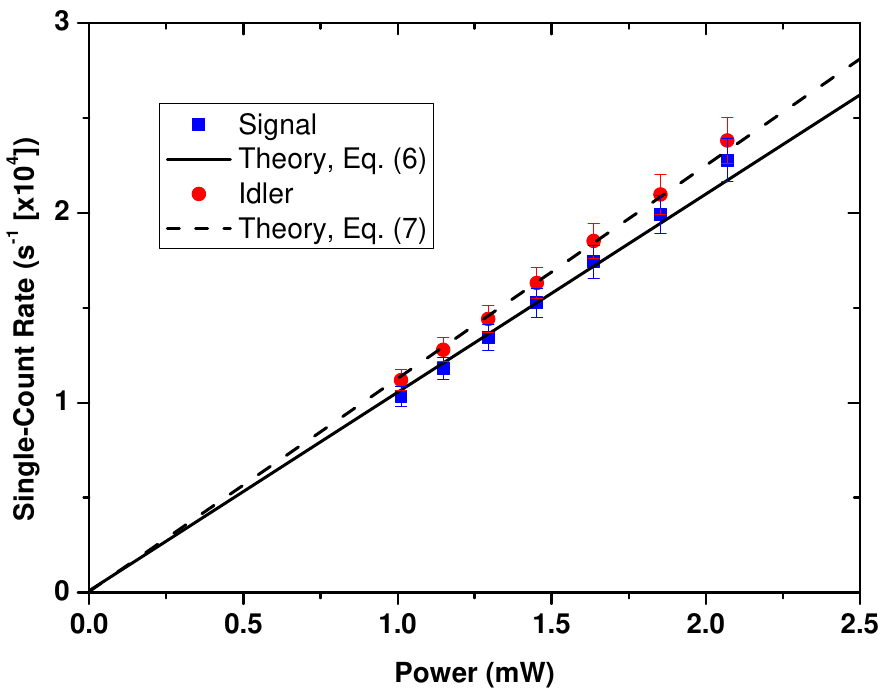}
\caption{Single-count rate of signal and idler photons, as a function of the fiber input pump power.}
\label{singles}
\end{figure}

The single-count rates of the signal and idler photons that are detected at the APDs, as a function of the fiber input pump power are presented in Fig.~\ref{singles}. As can be seen from the figure, there exists a linear relation between the single-count rates and the pump power, for both cases. The theoretical prediction is also plotted, using equations~\eqref{ns} and~\eqref{ni}. The parameters used in both equations were the following: $\eta_\textrm{s,i}\!=\!0.05$, $d_\textrm{s}\!=\!100$, and $d_\textrm{i}\!=\!50$. A good correspondence between theoretical and experimental results is verified. The difference between the count rates is due to the fact that the SpRS creates more noise photons on the idler (Stokes) than on the signal (anti-Stokes)~\cite{Lin.2007PhRvA..75b3803L}.
\begin{figure}[!b]
\centering
\includegraphics[width=8.5cm]{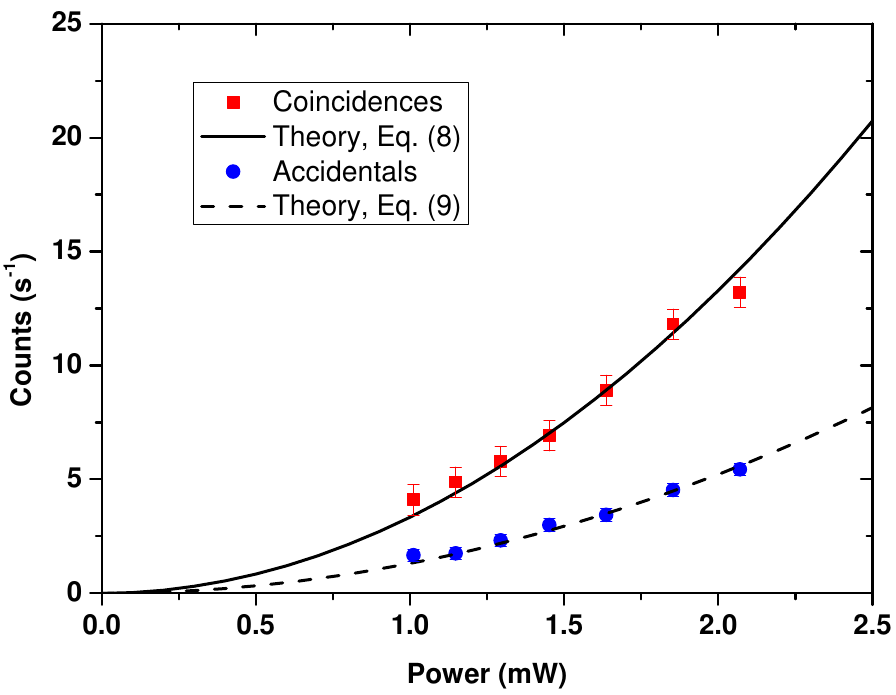}
\caption{Coincidence- and accidental-count rate as a function of the fiber input pump power.}
\label{coinc_acc}
\end{figure}

The coincidence- and accidental-count rates as a function of the fiber input pump power are presented in Fig.~\ref{coinc_acc}. The theoretical equations are also plotted, using equations~\eqref{nco} and~\eqref{nac}. The efficiency of the TTM was found to be $\eta_{\ms \textrm{$\alpha$}}~\approx~0.2$. A good agreement between theoretical and experimental data is also verified.

\begin{figure}[!t]
\centering
\includegraphics[width=8.5cm]{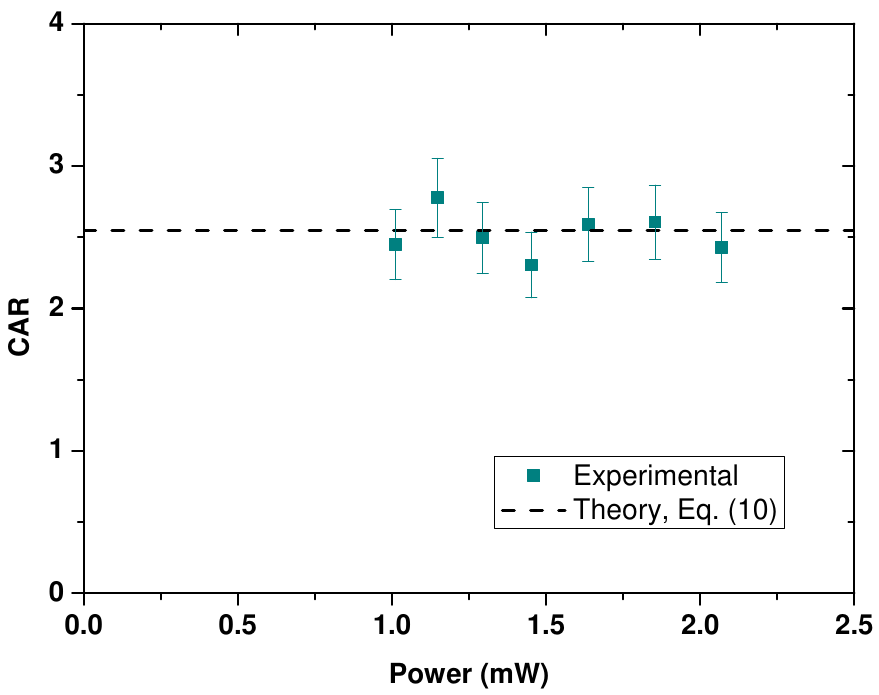}
\caption{CAR as a function of the fiber input pump power.}
\label{car}
\end{figure}

Finally, we analyze the quality of our source in terms of the CAR parameter. The experimental results and the theoretical prediction obtained from~\eqref{careq}, are plotted in Fig.~\ref{car}.
We have obtained a maximum CAR value of 2.77, that is verified for the input pump power of 1.1 mW. These results are comparable to the ones presented in other works~\cite{Zhou:09,Zhou:10}. Nevertheless, some improvements can be made in terms of the SpRS by cooling the fiber, as presented in~\cite{Takesue.2005OExpr..13.7832T}.
The reduced collection efficiency of the TTM also contributed significantly to the low CAR value, as its high time resolution window.

\section{Conclusion}
\label{c}

In conclusion, we have presented an experimental setup capable of producing correlated photon pairs through the SpFWM process in a HNLF, in the 1550~nm wavelength band. This fiber presents a nonlinear parameter higher than a DSF, which is traduced in several advantages. A smaller length and power can be used, and thus, leading to a lower rate of noise-photon generation through SpRS. We have calculated the photon-pair generation rate through SpFWM process and the noise-photon generation by SpRS, in the HNLF. The single-count rates at the signal and idler sides were also registered. A linear behavior was observed, increasing the counts with the power increase, as expected. Coincidence and accidental counts have been measured with a TTM, verifying a good relation between experimental and theoretical data. Finally, we have calculated the CAR, that presented a maximum value of 2.77, for an input pump power of 1.1~mW. The low collection efficiency and the large time window resolution of the TTM, were two parameters that contributed significantly to the reduced CAR value. Cooling the fiber also should lead to several improvements on the results. Nevertheless, the results obtained are in agreement with the ones presented in similar works.


\end{document}